\documentclass[intlimits,twoside,a4paper]{article}

\usepackage[cp1251]{inputenc}

\usepackage[eqsecnum]{cmpj3}

\usepackage{graphicx}
\usepackage{float}
\usepackage[caption = false]{subfig}
\newcommand{\kdp}{\mbox{KH$_2$PO$_4$} }

\issue{2022}{25}{4}{43709}
\doinumber{10.5488/CMP.25.43709}

\title[Shell-model and ab initio calculations in KDP]%
{Shell-model and first-principles calculations of vibrational, structural and ferroelectric
properties of \kdp%
}

\author[R. E. Mench\'on, F. Torresi, J. Lasave, S. Koval]{R. E. Mench\'on\orcid{0000-0002-6325-4230}, F. Torresi\orcid{0000-0002-4999-0530}, J. Lasave\orcid{0000-0003-1393-1114}, S. Koval\orcid{0000-0002-6925-4266}\thanks{Corresponding author: \email{koval@ifir-conicet.gov.ar}.}}
\address{
	Instituto de F\'{\i}sica Rosario, Universidad Nacional de Rosario and CONICET, 27 de Febrero 210 Bis, 2000 Rosario, Argentina
}

\Keywords{ferroelectrics, hydrogen bonds, phase transitions, shell model, density functional theory} 

\date{Received July 07, 2022, in final form August 12, 2022}

\begin{document}

\maketitle

\begin{abstract}
	We develop a shell model (SM) for
potassium dihydrogen phosphate (KDP) which is fitted to
{\it ab initio} (AI) results that include nonlocal van der Waals
corrections.  The SM is comprehensively tested
by comparing  results of structural, vibrational and ferroelectric properties
with AI and experimental data. The relaxed structural para\-meters  are in very 
good agreement with the AI results and the available experimental data.  The $\Gamma$-point phonons and the total phonon  densities of states (DOSs) in the ferroelectric and paraelectric phases calculated with the developed SM are in good overall 
agreement with the corresponding AI and experimental data. We also compute the effective Debye temperature as
a function of $T$ which shows good accordance with the corresponding AI and experimental results.
Classical molecular dynamics (MD) simulations obtained with the developed SM show a
FE--PE phase transition at $\approx 360$~K in remarkable agreement with {\it ab initio} MD calculations.
\printkeywords


\end{abstract}

\section{Introduction}

The prototype of the H-bonded ferroelectric (FE) compounds, \kdp or KDP, 
was extensively studied in the past due to its important technological applications as
well as for fundamental interest in its phenomenology~\cite{Lin77}.
In subsequent years after its discovery, KDP found extensive applications 
as electro-optical devices, as well as filter modulators
in the wide field of laser spectroscopy. Besides the research in technological aspects,
many efforts were also directed in the past to unveil the puzzle 
in the origin of the huge isotope effect that KDP manifests: the critical temperature $T_c$ 
for the paraelectric-ferroelectric (PE--FE) phase transition nearly doubles
when the system is deuterated~\cite{McM90,Nel91}. In spite of these efforts, the controversy
about whether tunneling~\cite{Bli60,Rei02} or geometrical effects~\cite{McM90,Nel91} 
are at the root of the giant isotope effect still remains.

This phenomenology is widely observed in the whole family of H-bonded ferroelectric compounds~\cite{Bli87,Kov05,Las07,Las11,Las16,Zac14,Zac15,Shc16}
including organic ferroelectrics that were recently discovered
which attract much attention because of their potential
for ecofriendly technological applications~\cite{Hor05,Hor10}.
\textit{Ab initio} (AI) calculations have shown that tunneling and geometrical effects
are involved in a self-consistent phenomenon
that greatly amplifies the effect of their correlation leading to the huge isotope effect observed~\cite{Kov02,Kov05}.

With the aim of studying the phase transition and the isotope effects in KDP, many models were
developed in the past~\cite{Bli60,Kob68,Tok87,Bli87,Sug91,Mer02}, e.g., the tunneling model and subsequent
modifications~\cite{Bli60,Kob68,Tok87,Bli87}. The  coupled proton-phonon model, which accounts for the
interaction between the proton collective mode exhibiting tunneling
and the optical lattice phonon of B$_2$ symmetry polarized along the $z$-axis, was invoked to explain the 
emergence of polarization in the $z$-direction~\cite{Kob68,Tok87}.
The lower-frequency mode arising from this interaction
is the FE soft mode that softens with temperature
when the critical temperature is approached.
Infrared measurements have also shown strong correlations
between these modes and an overdamped internal mode 
$\nu_4$ related to a quadrupole distortion of the phosphates~\cite{Sim88}.
Recent AI calculations have confirmed the strong couplings
of the FE soft mode with the lattice B$_2$ phonon and
the internal $\nu_4$ mode, as well as predicted important correlations with the phosphate libration mode 
of A$_1$ symmetry~\cite{Men18}. The latter could be relevant to the geometrical
effect observed because phosphate librations are associated with a change in the H-bond geometry.

Recently AI calculations were  successfully used to study phonons and vibrational properties in H-bonded FE compounds~\cite{Shc16,Men18}. However, generally, the AI approach is not computationally feasible
when dealing with  dynamical calculations in large-size systems to simulate 
the phase transition near $T_c$, where long-range effective interactions are essential. Even more demanding 
would be to simultaneously describe the quantum
nature of protons or deuterons, e.g., via path integral
simulations, although some AI calculations including nuclear
quantum effects were recently performed but with the focus on the PE phase properties
rather than on the phase transition itself~\cite{Sri11,Eng18}.
Moreover, the development of a SM  adjusted to AI results yielding reliable and transferable interatomic potential parameters,
especially for the H-bond unit, could be useful to undertake calculations on complex systems like some organic FE materials~\cite{Hor05}.

In spite of their
importance, lattice dynamics studies with atomistic models  in H-bonded
ferroelectrics like KDP were remarkably scarce in the past~\cite{Fuj70,Shc99,Shc06}. These
studies could be very useful to provide reliable interatomic potential
parameters in order to perform atomistic simulations for the study of the FE
phase transitions that these materials exhibit. On the other hand, due to the lack of confident microscopic information, 
many models developed in the past for these systems were validated \textit{a posteriori}
on the basis of their predictions. For example, indications of the existence
of tunneling were recently reported by Compton neutron scattering experiments~\cite{Rei02}, many
decades after the development of the tunneling model and its modifications.
With the advent of the electronic structure calculations based on the DFT theory, 
we have now the possibility of tuning the parameters of a model to reproduce
microscopic details of the system obtained from first principles.
Thus, it is desirable to develop an atomistic model adjusted to confident AI results.

A preliminary atomistic shell model (SM) for KDP~\cite{Las09,Kov10} was 
fitted to AI results obtained with the PBE-GGA functional~\cite{Kov02,Kov05}. 
A SM was chosen for the atomistic description of the system because
this kind of models was succesful in describing the large
oxygen polarization in related systems such as ferroelectric
perovskites~\cite{Sep05} or other oxide materials~\cite{Kov92,Kov96,Kov99,Gie06,Cas13}.
The work of reference~\cite{Las09} represents a first step towards the development of a reliable atomistic model necessary for 
the study of the phase transition with simulations of large-size systems.
However, the fit yielded a rather contracted geometry for the H-bonds and 
small values for the basal lattice parameters compared to the experiment. 
This is due to the fact that the model was fitted to the results
from PBE calculations  which, due to the neglect of the van der Waals (vdW) interactions,
underestimate the O--O distances and the energy-barriers for the polarization-inversion~\cite{Wik14,Las16,Men18}.
For instance, the energy barriers  obtained at the PBE level~\cite{Las09} are of the order of $\approx 4$
to 5 times and $\approx 10$ times smaller than those calculated with the non-local vdW-DF approach and the
M\"{o}ller-Plesset second order perturbation theory (MP2), respectively~\cite{Men18}.
Similarly, the PBE results for the O--O distance and the parameter $\delta$, which is twice the proton distance
to the middle of the O--H--O bond, appears $\approx 4$\% and $\approx 50$\% smaller, respectively, than 
the corresponding values of both, the vdW-DF and MP2 methods~\cite{Men18}.
We remark here that the approximate inclusion of nuclear quantum effects~\cite{Las16,Men18} for
the PBE results lead to an important disagreement with the experiment 
for the last mentioned magnitudes, especially the equilibrium proton
distance $\delta$ which appears to be too short: 0.015~\AA{} compared to
0.385~\AA{} for the experimental value~\cite{Men18}. On the other hand,
nuclear quantum corrections (NQC) applied to the vdW-DF case
lead to a value of $\delta=0.355$~\AA{}, in excellent agreement with the experiment~\cite{Men18}.
The high-energy phonons related to the H-bond dynamics have also much smaller frequencies
in the PBE scheme than those obtained with nonlocal vdW approaches such as the vdW-DF method~\cite{Fin16,Men18}.
We conclude that the SM developed in reference~\cite{Las09}
would not be suitable for describing correctly  the phase transition due to the 
resulting contracted H-bond geometry and the small global proton-transfer 
energy barriers. 

In this work, we have developed a new SM based on
interatomic potentials by adjusting its parameters to old and new AI structural data that include vdW corrections using the so-called
vdW-DF approach, which had the best performance
among different AI methods to reproduce experimental structural and vibrational properties of KDP and DKDP~\cite{Men18}.
The developed SM, together with new~AI calculations, were used to study different ferroelectric, structural and vibrational properties.
The~SM calculations included full structural relaxations, zone-center phonons, phonon densities of states, the effective Debye temperature
and a ``classical-nuclei'' SM molecular dynamics study of the phase transition which were compared to AI molecular dynamics simulations also performed here.
In this paper we have also determined new full structural AI relaxations in the PE and FE phases, and computed the~AI total phonon density of states in the PE phase. These data together with previous AI data from reference~\cite{Men18} and avai\-lable experimental results were used to extensively test the results of the new model developed.
The~SM results for the effective Debye temperature show good agreement compared to AI~\cite{Men18} and experimental data. Moreover, we have found a very good agreement in the results of the~SM and AI molecular dynamics simulations performed for the study of the phase transition in deuterated~KDP~(DKDP).

This paper is organized as follows: in section~\ref{sec:2} we give details
about the SM developed and the AI and SM calculations performed.
The results are presented in section~\ref{sec:3}. In the first subsection~\ref{sec:3_1} we analyze the structures obtained with the present SM
for the PE and FE phases and compare them 
with AI~(vdW-DF) results as well as with the available experimental data. 
The SM calculations of the phonons and related  
properties are presented in section~\ref{sec:3_2}, which are in turn
divided into two subsections. In the first subsection~\ref{sec:3_2_1}
we report some of the SM phonons at the Brillouin zone center
and compare them with AI and experimental data.
Besides, we also present a comparison of the total DOS in the FE and PE phases
with the corresponding AI results.
In the second subsection~\ref{sec:3_2_2}, we analyze the
SM results for the  effective
Debye temperature and compare them with the
corresponding AI and experimental data.
In section~\ref{sec:3_3} we present  molecular dynamics simulations for the developed SM and compare them with AI results.
Finally, a summary is presented in section~\ref{sec:4}.

\section{Shell model and \textit{ab initio} methods. Calculation details}
\label{sec:2}

The PE phase of KDP has a body-centered tetragonal
$bct$ structure with shorter lattice constant along the tetragonal axis.
The space group is I$\bar{4}$2d or D$_{2d}^{12}$.
In the FE phase, the crystal is 0.8\% shear distorted along the [110] direction and 
becomes orthorhombic with space group Fdd2 or C$_{2v}^{19}$.
The primitive cell in both phases contains 16 atoms (two formula units).

In the PE phase of KDP, the hydrogens have two equilibrium
positions equidistant to the middle of the O--H--O bond and separated 
by a distance $\delta$. Both positions are occupied by these
protons with equal probability in this phase, and hence
the averaged proton position is $\left \langle x \right \rangle = \langle \delta \rangle /2 = 0$.
Therefore, in the AI and SM simulations for this phase, we performed
full structural optimizations with the H atoms exactly at the middle of the O--H--O bonds~\cite{Shc16}, which remained in their positions after relaxations due to symmetry reasons. This hypothetical phase at 0~K is used to  keep the macroscopic center inversion symmetry of I$\bar{4}$2d phase. For this phase, the phonon calculations give three unstable modes of imaginary frequency, one of which is shown in table~\ref{phonon1}.
In the case of the FE phase, the protons are originally slightly displaced from the middle of the O--H--O bonds following 
the pattern of the FE mode. After that, 
the full structural SM and AI relaxations lead the system in each case to the ordered FE phase.
In the SM and AI optimized-cell structural relaxations for both phases, the cell plus atomic structural parameters were allowed to relax at zero pressure.

Our starting point for the SM is the one developed in reference~\cite{Las09}.
It contains ionic polarizabilities for the P and O ions
which are described by an electronic shell with charge $Y$ harmonically
coupled with a spring $k_{sc}$ to the core.
In addition, we consider short-range shell-shell repulsive
interactions arising from the wavefunction overlap between neighboring ions
and long-range Coulomb interactions between cores and shells of every ion. The short-range interactions are
of the Born--Mayer type, $A \re^{-r/\rho}$, for the P--O and O--H bonds as well as
of the Buckingham type, $A \re^{-r/\rho}-C/r^6$, for the K--O bonds~\cite{Kov92,Kov96,Cas13}.
Three-body angular
potentials of the form $ (1/2)k(\theta-\theta_0)^2$ 
for the covalent O--P--O and P--O--H bonds~\cite{Cas13}
and a three-body potential represented by the term 
$ (D/r^{12}-B/r^{10})\cos^4(\theta-180^{\circ}) $ for the O--H--O bond
are also included~\cite{Las09}. Considering the total charge
$Z$ of the ions and using the condition of
charge neutrality, the model finally contains 20 adjustable parameters.

\begin{table*}
	\caption{Shell model parameters for the present work.}
	\label{table_modelparams}
	\label{parameters}
	\begin{center} 
		\scalebox{0.95}{\begin{tabular}{|c c c c c c c|}
				\hline
				\hline
				{ } & {H$_{\rm {core}}$} & {O$_{\rm {core}}$} & {P$_{\rm {core}}$} & {K$_{\rm {core}}$} & {O$_{\rm {shell}}$} & {P$_{\rm {shell}}$}\\
				\hline
				{\bf charge [e]} & {0.65} & {1.10} & {4.80} & {0.60} & {$-2.40$} & {$-1.50$}\\
				\hline
				\hline
		\end{tabular}}
		$$\quad$$
		\scalebox{0.85}{\begin{tabular}{|c c c c||c c c|}
				\hline
				\hline
				\multicolumn{4}{|c||}{\bf Two-Body Potentials} & \multicolumn{3}{|c|}{\bf Three-Body Potentials}\\
				\hline
				{\bf    harmonic    } & {    $k_{sc}$ [eV~\AA{}$^{-2}$]    } & {        } & {        } & {\bf    angular harmonic    } & {    $k$ [eV rad$^{-2}$]    } & {    $\theta_0$ [\textdegree]    }\\
				\hline
				{    O$_{\rm {core}}$--O$_{\rm {shell}}$    } & {    68.0    } & {        } & {        } & {    O$_{\rm {core}}$--P$_{\rm {core}}$--O$_{\rm {core}}$    } & {    18.00    } & {    109.0    }\\
				{    P$_{\rm {core}}$--P$_{\rm {shell}}$    } & {    1950.0    } & {        } & {        } & {    H$_{\rm {core}}$--O$_{\rm {core}}$--P$_{\rm {core}}$    } & {    2.75    } & {    115.8    }\\
				\hline                                                        
				{\bf    Buckingham    } & {    $A$ [eV]    } & {    $\rho$ [\AA{}]    } & {    $C$ [eV~\AA{}$^6$]    } & {\bf    hydrogen bond 12-10-4    } & {    $D$ [eV~\AA{}$^{12}$]    } & {    $B$ [eV~\AA{}$^{10}$]    }\\
				\hline
				{    H$_{\rm {core}}$--O$_{\rm {shell}}$    } & {    184.5    } & {    0.21    } & {    0.0    } & {    O$_{\rm {core}}$--H$_{\rm {core}}$--O$_{\rm {core}}$    } & {    20478.0    } & {    1750.0    }\\
				{    O$_{\rm {shell}}$--P$_{\rm {shell}}$    } & {    805.0    } & {    0.30    } & {    0.0    } & {        } & {        } & {        }\\
				{    O$_{\rm {shell}}$--K$_{\rm {core}}$    } & {    5175.0    } & {    0.32    } & {    507.5    } & {        } & {        } & {        }\\
				\hline
				\hline
		\end{tabular}}
		
	\end{center}
\end{table*}

The SM simulations were carried out using the GULP
code~\cite{Gal97} which can perform structure optimizations at zero temperature
as well as phonon calculations (phonon frequencies and eigenvectors
at the Brillouin zone center, phonon dispersion curves and density of states, etc.) and classical molecular dynamics 
simulations.

The AI calculations in the PE and FE phases of KDP were performed
with the VASP code using projector augmented wave (PAW) all-electron potentials~\cite{Kre96a,Kre96b}.
The plane-wave basis was
expanded to an energy cutoff of 750~eV.
In the calculations we use an automatic Monkhorst-Pack $5\times5\times5$ grid 
sampling of the electronic Brillouin zone, which
proved sufficient to achieve converged results.
The calculations were carried out using 
the AI scheme vdW-DF that includes nonlocal
van der Waals dispersion corrections~\cite{Dio04,Kli11,Rom09}.
The geometry optimizations were performed until the
forces on every atom were smaller than 5~meV/\AA{}.

Shell model molecular dynamics (SMMD) and 
{\it ab initio} molecular dynamics (AIMD) simulations were performed in the NVT ensemble
with the GULP and VASP programs, respectively, using
a Nos\'e-Hoover thermostat to control the temperature. 
The Newton's equations of motion were integrated using the Verlet's algorithm with a typical time step of 0.3~fs~(0.2~fs) in the AIMD~(SMMD) simulations for an accurate integration of the electronic degrees of freedom in H-bonded systems. The AIMD~(SMMD) calculations were carried out for a series of temperatures from 200 to 500~K using supercells of 128~(256) atoms subjected to periodic boundary conditions. At each temperature, a well equilibrated configuration is achieved after running at least 5~ps. 
Then, the system evolves further at least 5 and 15~ps to calculate the thermodynamic averages in the AIMD and SMMD
calculations, respectively.

The AIMD calculations were performed with a high-accuracy cutoff energy of 600~eV and $\Gamma$-point Brillouin zone sampling, and include 
nonlocal dispersion corrections at the vdW-DF level. 
It is worth mentioning that in order to ensure that the energy is well converged we have verified that the drift produced by the Nos\'e-Hoover dynamics is less than 1~meV/atom within a time step of 1~ps~\cite{Zha14}.

The lattice parameters were fixed in the simulations for DKDP to those
corresponding to the expe\-rimental values in the PE phase measured
at $T_c + 5$~K~$\approx 234$~K~\cite{Nelmes87}. Accordingly,
the ordered phase at low temperature arises in the calculations at fixed cell (NVT ensemble) as a FE phase with tetragonal lattice, which is slightly distorted with respect to the orthorhombic lattice~\cite{Kov01}. This
strategy enables the study of the phase transition using a supercell that
conserves volume.
We verified that all AIMD and SMMD simulations yield to average equilibrium structures which are stable up to the largest temperature considered $T=500$~K.

The total {\it ab initio} phonon density of states (PDOS) was derived from the phonon calculations using the finite difference (FD) method as implemented in the PHONOPY code~\cite{Tog15,Kre95,Men18}. Here, in order to compute the atomic forces for the different configurations generated by the PHONOPY FD method
we used the VASP program with the vdW-DF method to account for nonlocal dispersion effects. In this case, the energy cutoff for the plane-wave basis was set to 450~eV, and a $4\times4\times4$ grid for the Brillouin zone sampling was used. The atomic distortions were performed in a $2\times2\times2$ supercell with 128 atoms. In this case we used a tighter tolerance in the forces of 0.5~meV/\AA{} in order to achieve convergence in the phonon dispersion results. Once the phonon dispersion relations were obtained, the PHONOPY code allowed us to calculate the total PDOS by a Brillouin zone integration. 
 \clearpage
\section{Results and discussion}
\label{sec:3}

\subsection{Structural optimizations}
\label{sec:3_1}

The model parameters of the SM were initially taken from reference~\cite{Las09} and were further adjusted in order to reproduce the 
AI~(vdW-DF) results for the relaxed internal structure 
and lattice constants in the FE and PE phases.
Special emphasis was made on reproducing
the O--O distance and the parameter $\delta$
for the H-bonds, which were mainly controlled
by adjusting the three-body potential for the O--H--O bond
and the Born--Mayer O--H potential.
Once the set of parameters satisfactorily reproduced the structure of both phases, we further
refined the model to adjust better the Brillouin zone-center 
($\Gamma$-point) phonons to the AI data.
In doing this, due to the complexity of the system,
special care was taken to adjust various properties
at the same time without spoiling the whole fit.

In table~\ref{table_structure}, we present results of the structural parameters
obtained with full AI and SM relaxations (atomic plus cell optimizations) 
in the PE and FE phases.  These results are compared to the corresponding AI values from reference~\cite{Men18}. 
A good overall agreement between the AI~(vdW-DF) and SM structural parameters results for both phases
is observed in table~\ref{table_structure}.
Moreover, the comparison of these results with the available experimental data is also satisfactory,
with relative percentage differences of the order or less than  3\% in most of the cases.
The underestimation in the O--O distance in the PE phase can be 
mainly attributed to the static optimization for centered protons in the H-bonds~\cite{Kov02,Kov05}, while in the actual PE phase
observed at finite temperature, the protons
are delocalized over two symmetric, off-center positions
along the bond~\cite{McM90,Nel91}. 
Notice that the theoretical calculations correspond to classical (infinite mass) nuclei.
Actually, H-bond parameters such as the O--O, O--H and $\delta$ distances are especially
affected by the quantum dynamics of the proton~\cite{Las16,Men18}. The inclusion of these dynamics
approximately by means of PIMC calculations, leads to NQC that improve
the correspondence of the AI results for these parameters~\cite{Men18} with the experiment.
Notice that the present SM calculations
overestimate the relaxed H-bond geometry in the FE phase in comparison with the experiment, e.g.,
the O--O distance and the parameter $\delta$
(see table~\ref{table_structure}). Therefore, due to the fact
that NQC produce a contraction in the H-bonds~\cite{Men18},
we speculate that if NQC were applied to the SM calculations,
the agreement with the experiment for the H-bond geometry could be improved.

\begin{table*}[h]
	\caption{SM and AI~(vdW-DF) results of the lattice and internal structure parameters for the FE and PE phases of KDP. Also shown is the experimental data of reference~\cite{Nelmes87}.
		Distances in angstroms and angles in degrees.  We also show in parenthesis the relative percentage differences with available experimental data.}
	\label{table_structure}
	\label{validation}
	\begin{center} 
			\scalebox{0.89}{\begin{tabular}{|c | c c c | c c c|}
					\hline
					\hline
					{\bf Structural } & \multicolumn{3}{|c|}{\bf FE structure} & \multicolumn{3}{|c|}{\bf PE structure }\\
					{\bf Parameters } & {\it Expt.}~\cite{Nelmes87} & { vdW-DF } & { SM } & {\it Expt.}~\cite{Nelmes87} & { vdW-DF } & { SM }\\
					{ } & { } & { $\textrm{\footnotesize optimized cell}$  } & { $\textrm{\footnotesize optimized cell}$  } & { } & { $\textrm{\footnotesize optimized cell}$  } & { $\textrm{\footnotesize optimized cell}$  }\\
					\hline                  
					\hline              
					{ $a$ }                         & { 10.546 }  & { 10.836 (2.8) } & { 10.893 (3.3) } & { 7.426 } & { 7.531 (1.4)} & { 7.258 (2.3)}\\
					{ $b$ }                         & { 10.466 }  & { 10.741 (2.6)} & { 10.099 (3.5) } & {  }      & {  }      & {  }\\
					{ $c$ }                         & { 6.927 }   & { 7.119 (2.8)} & { 6.925 (0.1) } & { 6.931  }   & { 7.096 (2.4) } & { 6.933 (0.3) }\\
					\hline                  
					{ $d$ (O$\dots $O) }             & { 2.497 } & { 2.612 (4.6)} & { 2.585 (3.5) } & { 2.483 } & { 2.430 (2.1) } & { 2.429 (2.2) }\\
					{ $d$ (O$_2$--H) }                   & { 1.056 } & { 1.022 (3.2) } & { 0.949 (10) } & { 1.071 } & { 1.215 (13)} & { 1.217 (13) }\\
					{ $d$ (H$\dots$O$_1$) }            & { 1.441 } & { 1.590 (10)} & { 1.637 (14)} & { 1.412 } & { 1.215 (14) } & { 1.217 (14)}\\
					{ $\delta$ }                    & { 0.385 } & { 0.568 (48)} & { 0.687 (78)} & {  }      & {  }      & {  }\\
					{ $\angle$ (O$_2$--H$\dots$O$_1$) } & { 179.4 } & { 179.8 (0.2)} & { 177.7 (0.9)} & { 178.2 } & { 178.5 (0.2)} & { 173.0 (2.9)}\\
					{ $d$ (P$\dots$K)}             & { $-$ }     & { 3.402 } & { 3.384 } & { $-$ }     & { 3.548 } & { 3.467 }\\
					{ $d$ (K$\dots$P)}             & { $-$ }     & { 3.717 } & { 3.542 } & { $-$ }     & { 3.548 } & { 3.467 }\\
					{ $d$ (P--O$_1$) }                   & { 1.516 } & { 1.524 (0.5)} & { 1.527 (0.7)} & {  }      & {  }      & {  }\\
					{ $d$ (P--O$_2$) }                   & { 1.572 } & { 1.617 (2.9)} & { 1.692 (7.6)} & { 1.543 } & { 1.564 (1.4)} & { 1.595 (3.4)}\\
					{ $d$ (K$\dots$O$_1$) (nn) }       & { 2.785 } & { 2.854 (2.5)} & { 2.707 (2.8)} & {  }      & {  }      & {  }\\
					{ $d$ (K$\dots$O$_2$) (nn) }       & { 2.825 } & { 2.908 (2.9)} & { 2.730 (3.4)} & { 2.809 } & { 2.888 (2.8)} & { 2.672 (4.9)}\\
					{ $d$ (K$\dots$O$_1$) (nnn) }      & { 2.847 } & { 2.914 (2.4)} & { 2.811 (1.3)} & {  }      & {  }      & {  }\\
					{ $d$ (K$\dots$O$_2$) (nnn) }      & { 2.914 } & { 3.011 (3.3)} & { 2.891 (0.8)} & { 2.881 } & { 2.952 (2.5)} & { 2.859 (0.8)}\\
					\hline            
					\hline
			\end{tabular}}
			
		\end{center}
	\end{table*}

\subsection{Vibrational properties}
\label{sec:3_2}

\subsubsection{$\Gamma$-point phonons and density of states}
\label{sec:3_2_1}

We have checked the capability of the developed SM to reproduce vibrational
properties of the system obtained from first principles calculations. 
To this end, we compare some of the SM Brillouin zone-center phonons with the
corresponding AI~(vdW-DF) and experimental data in table~\ref{phonon1}.
The table is organized in view of the correspondence between
representations in the PE and FE phases.~\cite{Men18}
Therefore, frequencies in the same row correspond
to qualitatively similar patterns of motion for both phases,
although the correspondence should be taken with caution
because in some cases there were ambiguities in the assignement.
In table~\ref{phonon1}, we show the SM results for the zone-center
phonons of symmetry  A$_1$, A$_2$, B$_1$, and~B$_2$ in the PE phase,
which are divided into the two subspaces, (${\rm A}_1 + {\rm B}_2$) and
(${\rm A}_2 + {\rm B}_1$). In the right-hand panel we report the 
corresponding SM frequencies of the A$_1$ and A$_2$ modes in the FE phase.

\begin{table*}[h]
	\caption{SM results of the $\Gamma$-point phonons of symmetries   $A_1$, $B_2$, $A_2$, $B_1$ (PE~phase) and $A_1$, $A_2$ (FE~phase) for KDP. 
		Also shown are the AI~(VASP) results obtained with the exchange-correlation functional vdW-DF from reference~\cite{Men18} and the experimental results of references~\cite{Coi71,She71,Ser88} for the PE phase and references~\cite{Coi71,Tom83b,Sim88} for the FE phase. The phonon frequencies are shown in cm$^{-1}$. 
		According to the calculated eigenvectors, the following classification is shown in the table: 
		ferroelectric unstable mode with imaginary frequency (FE), external traslational (ET), external rotational (ER),
		internal molecular P--O bending (IMB), internal molecular P--O stretching (IMS), O--H$\dots$O bending (HB), 
		and O--H$\dots$O stretching (HS).
	} 
	\label{phonon1}
	\begin{center} 
		\scalebox{0.56}{\begin{tabular}{|c c c c c c c|c c c c c c c|}
				\hline
				\hline
				\multicolumn{7}{|c|}{\bf PE structure} & \multicolumn{7}{c|}{\bf FE structure}\\
				\multicolumn{1}{|c}{Sym.} & {SM} & {vdW-DF~\cite{Men18} } & \multicolumn{1}{c}{{\it Expt.}~\cite{Coi71}} & \multicolumn{1}{c}{{\it Expt.}~\cite{She71}} & \multicolumn{1}{c}{{\it Expt.}~\cite{Ser88}} &\multicolumn{1}{c|}{Class.} & \multicolumn{1}{c}{ Sym. }  & {SM} & { vdW-DF~\cite{Men18} } & \multicolumn{1}{c}{{\it Expt.}~\cite{Coi71}} & \multicolumn{1}{c}{{\it Expt.}~\cite{Tom83b}} & \multicolumn{1}{c}{{\it Expt.}~\cite{Sim88}} & \multicolumn{1}{c|}{Class.}\\
				\hline	
				\hline								
				
				{	B$_2$	} & {	1096i	} & {	819i	} & {		} & {		} & {		} & {	FE, HS + IMB	} & {	A$_1$	} & {	2528	} & {	2600	} & {		} & {		} & {		} & {	HS	}\\
				{	B$_2$	} & {	226	} & {	189	} & {	180	} & {	180	} & {	179.5	} & {	ET	} & {	A$_1$	} & {	227	} & {	185	} & {	185	} & {	142	} & {	209	} & {	ET	}\\
				{	A$_1$	} & {	648	} & {	543	} & {		} & {		} & {	520.3	} & {	IMB	} & {	A$_1$	} & {	274	} & {	257	} & {		} & {	283	} & {		} & {	IMB	}\\
				{	A$_1$	} & {	422	} & {	295	} & {	360	} & {	363	} & {	363.9	} & {	IMB	} & {	A$_1$	} & {	528	} & {	345	} & {	346	} & {	369	} & {		} & {	IMB	}\\
				{	B$_2$	} & {	610	} & {	391	} & {	395	} & {	393	} & {	394.0	} & {	IMB	} & {	A$_1$	} & {	595	} & {	383	} & {	394	} & {	393	} & {		} & {	IMB	}\\
				{	B$_2$	} & {	593	} & {	533	} & {		} & {		} & {		} & {	IMB	} & {	A$_1$	} & {	623	} & {	488	} & {	515	} & {		} & {		} & {	IMB	}\\
				{	A$_1$	} & {	784	} & {	917	} & {	915	} & {		} & {	916.9	} & {	IMS + HS	} & {	A$_1$	} & {	753	} & {	833	} & {		} & {		} & {	859	} & {	IMS + HB	}\\
				{	B$_2$	} & {	988	} & {	1145	} & {		} & {		} & {		} & {	IMS + HB	} & {	A$_1$	} & {	868	} & {	941	} & {	910	} & {		} & {		} & {	IMS + HB	}\\
				{	B$_2$	} & {	1162	} & {	1273	} & {		} & {		} & {		} & {	HB	} & {	A$_1$	} & {	1013	} & {	1002	} & {	1035	} & {		} & {	1048	} & {	HB + IMS	}\\
				{	A$_1$	} & {	1372	} & {	1340	} & {		} & {		} & {		} & {	HB	} & {	A$_1$	} & {	1300	} & {	1282	} & {		} & {		} & {		} & {	HB	}\\
				\hline
				{	B$_1$	} & {	94	} & {	117	} & {		} & {		} & {		} & {	ET	} & {	A$_2$	} & {	88	} & {	120	} & {		} & {		} & {		} & {	ET	}\\
				{	B$_1$	} & {	215	} & {	174	} & {	152	} & {	151	} & {	155.8	} & {	ET	} & {	A$_2$	} & {	216	} & {	179	} & {	154	} & {	159	} & {		} & {	ET	}\\
				{	A$_2$	} & {	103	} & {	224	} & {		} & {		} & {		} & {	ER	} & {	A$_2$	} & {	166	} & {	221	} & {	206	} & {	211	} & {		} & {	ER	}\\
				{	A$_2$	} & {	341	} & {	347	} & {		} & {		} & {		} & {	IMB	} & {	A$_2$	} & {	625	} & {	354	} & {		} & {		} & {		} & {	IMB	}\\
				{	B$_1$	} & {	584	} & {	475	} & {	475	} & {	470	} & {	474.5	} & {	IMB	} & {	A$_2$	} & {	550	} & {	467	} & {	485	} & {	483	} & {		} & {	IMB	}\\
				{	B$_1$	} & {	773	} & {	680	} & {		} & {		} & {	564.1	} & {	IMB + HS	} & {	A$_2$	} & {	672	} & {	529	} & {		} & {		} & {		} & {	IMB + HS	}\\
				{	A$_2$	} & {	641	} & {	826	} & {		} & {		} & {		} & {	IMS + HS	} & {	A$_2$	} & {	751	} & {	804	} & {		} & {		} & {		} & {	IMS + HB	}\\
				{	B$_1$	} & {	975	} & {	1021	} & {		} & {		} & {		} & {	IMS + HS	} & {	A$_2$	} & {	1028	} & {	1027	} & {	1008	} & {		} & {		} & {	IMS + HB	}\\
				{	A$_2$	} & {	742	} & {	1104	} & {		} & {		} & {		} & {	HS + HB	} & {	A$_2$	} & {	2580	} & {	2719	} & {		} & {		} & {		} & {	HS	}\\
				{	B$_1$	} & {	1371	} & {	1349	} & {		} & {		} & {		} & {	HB	} & {	A$_2$	} & {	1301	} & {	1282	} & {		} & {		} & {		} & {	HB	}\\
				{	A$_2$	} & {	1000	} & {	1277	} & {		} & {		} & {		} & {	HB	} & {	A$_2$	} & {	820	} & {	957	} & {		} & {		} & {		} & {	HB	}\\
				
				\hline
				\hline
		\end{tabular}}
		
	\end{center}
\end{table*}

This classification can be applied to the $\Gamma$-point phonon results
as well as to the phonon bands obtained from the phonon density of states (DOS) results
which are shown below.

The SM results for the chosen Brillouin zone-center  phonons are compared in table~\ref{phonon1} with the AI~(vdW-DF) results and with the available experimental data. A good overall agreement is observed between the SM frequencies and the 
corresponding AI~\cite{Men18} and experimental data, as well as with other AI phonon calculations for KDP~\cite{Shc21}. Despite the fact that we are analysing a hypothetical PE phase at 0~K where the protons are fixed centered in the H-bonds, and that the low frequency phonons and modes involving hydrogen atoms  may be affected by the three unstable modes arising from the used approximation, it is instructive to qualitatively compare the changes observed in the phonon spectrum between both PE and FE phases. For instance, one of the unstable modes in the PE phase of B$_2$ symmetry, which is identified by its imaginary frequency in table~\ref{phonon1}, corresponds to a similar one found with the AI calculations. Such B$_2$ mode is
the FE soft mode in the PE phase, which becomes stable in the FE phase as a high-frequency HS phonon (see table~\ref{phonon1}).

When the phase changes from PE to FE some modes 
soften while others stiffen and, with a few exceptions, the SM phonons follow the same trend as the AI phonons. In particular, 
one of the most significant frequency changes between both phases
is observed for the PO$_4$-rotational A$_1$ mode obtained at 648~cm$^{-1}$ in the PE phase for the SM calculation, which softens to 274~cm$^{-1}$ in the FE phase
(see table~\ref{phonon1}).
Similarly, this mode softens strongly in the AI calculation.
It is demostrated that this phonon couples significantly with the
FE soft mode and that this interaction becomes stronger as the
amplitude of the latter is increased~\cite{Men18}.
The FE soft-mode picture  
for the phase transition in H-bonded ferroelectrics
is consistent with recent NMR experiments which show
the importance of a displacive component near the
critical temperature~\cite{Kwe17,Dal98}.

\begin{figure*}[ht]
\centerline{\includegraphics[scale=0.30]{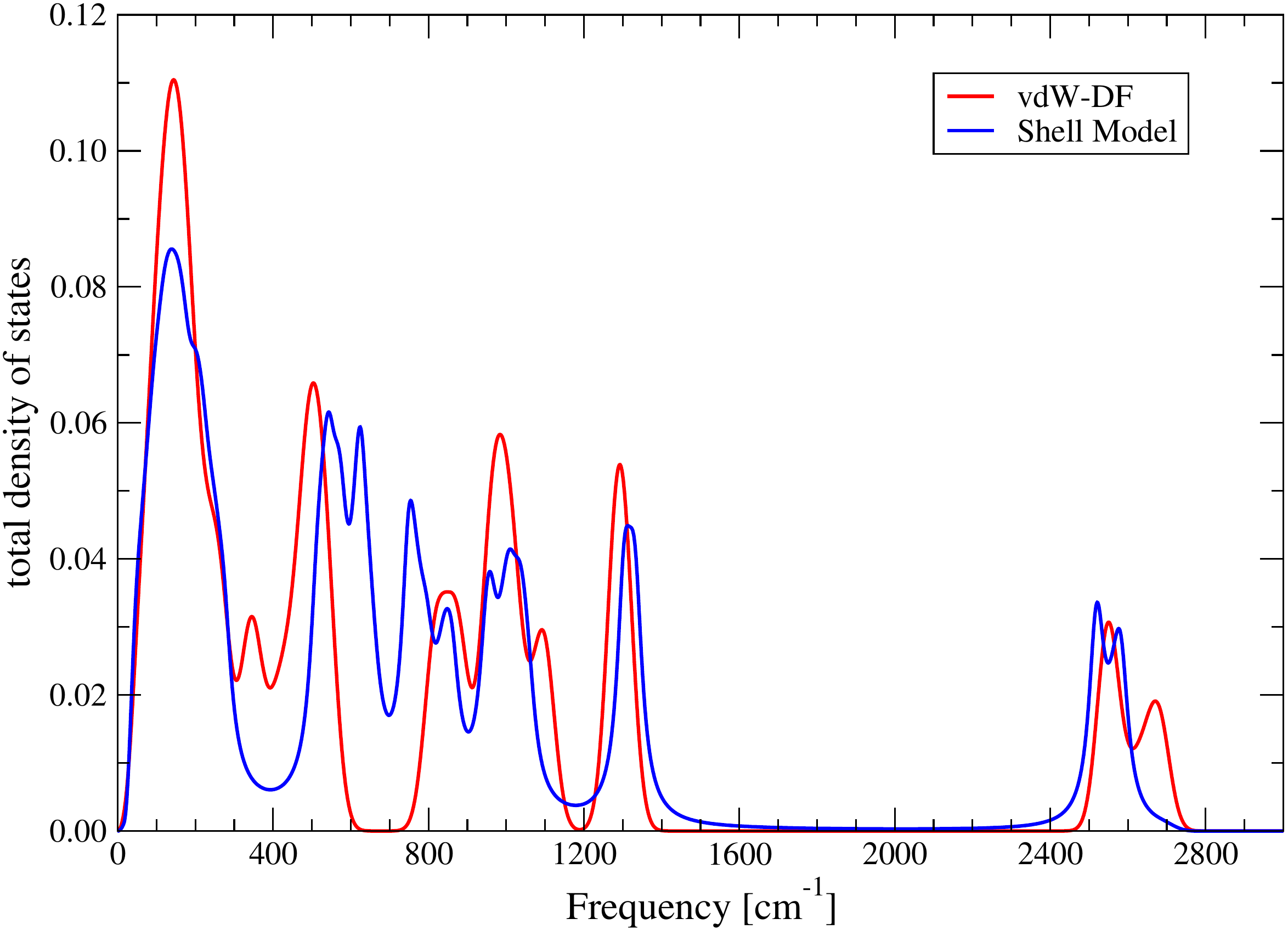} 
\hskip 0.2cm
\includegraphics[scale=0.30]{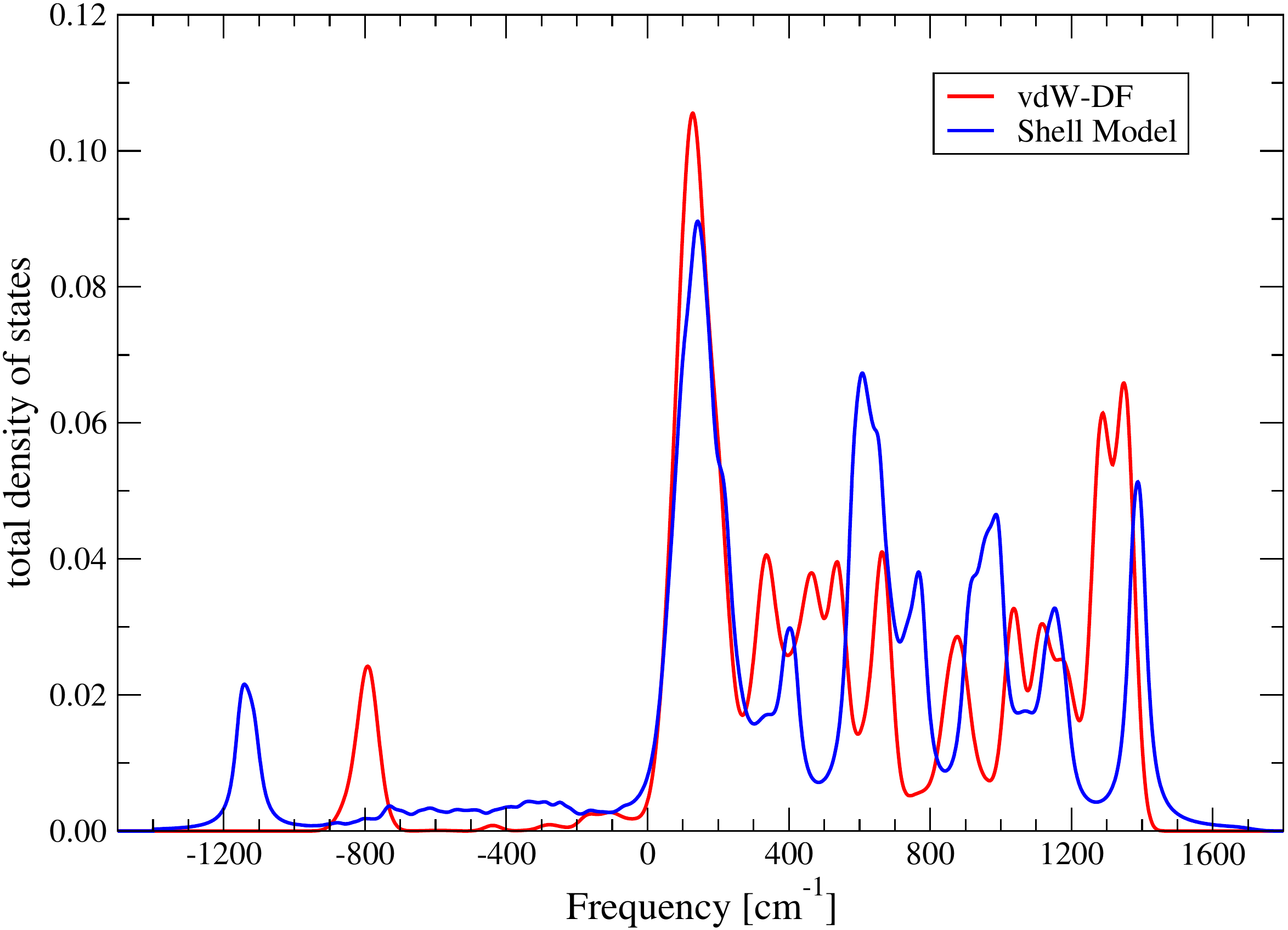}}
	\caption{(Colour online) Left-hand panel: SM (blue solid line) and AI (red solid line) 
		results of the total phonon density of states in the FE phase of KDP. 
		Right-hand panel: {\it idem} for the PE phase. The AI PDOS in the FE phase are results from reference~\cite{Men18}. }
	\label{fe_pe_dos.fig}
\end{figure*}

The total SM-DOSs in the FE and PE phases are plotted 
in figure~\ref{fe_pe_dos.fig}
and compared to the corresponding AI~(vdW-DF) results, showing
a general good agreement between both calculations for both phases.
Notice that the SM and AI~(vdW-DF) unstable phonon 
bands observed in the PE phase at imaginary frequencies (plotted
as negative frequencies in the right-hand panel of figure~\ref{fe_pe_dos.fig}), are related to the unstable FE  zone-center phonon reported 
in table~\ref{phonon1}.

From these plots and the classification of the phononic displacement patterns made in table~\ref{phonon1}, we can identify the different types of bands present in the spectra for the FE phase. The highest-frequency band corresponds to the HS band
centered at $\approx 2550$~cm$^{-1}$ for the SM calculation which is in very good agreement
with the AI result for the HS band centered at $\approx 2600$~cm$^{-1}$ as shown in the
left-hand panel of figure~\ref{fe_pe_dos.fig}. Particularly, the bimodal shape of this band and the
spectral weight are correctly described by the SM calculation, although the position of the highest-frequency peak of the
SM bimodal distribution is a bit shifted towards lower frequencies with respect to the AI result.

The SM result for the HB band is also in remarkable agreement with the corresponding AI result, displaying similar peak positions ($\approx 1300$~cm$^{-1}$) and band extensions 
in both calculations (see figure~\ref{fe_pe_dos.fig}).
The large gap between the HB and HS bands observed in the FE phase in the AI calculation 
is also qualitatively reproduced by the SM results as shown in figure~\ref{fe_pe_dos.fig}. 
 We also find that the SM DOS result is in qualitative agreement with another AI calculation for the FE phase~\cite{Fin16}. 

The SM lowest-frequency band in the FE phase centered at $\approx~150$~cm$^{-1}$ is also in very good
agreement with the corresponding AI data (see figure~\ref{fe_pe_dos.fig}).
This is a lattice external (ET+ER) band that extends up to~$\approx 300$~cm$^{-1}$, 
which is of K and O character mainly. 

The intermediate frequency region of the DOSs spectra for the FE phase consists
of two bands of internal molecular (IM) phonons involving primarily the
PO$_4$ tetrahedron. The high-frequency band extends from 
$\approx 700$ to $\approx 1200$~cm$^{-1}$ in the SM calculation
in close agreement with the AI result~\cite{Men18} (see figure~\ref{fe_pe_dos.fig}), 
and consists of the modes involving mainly the stretching of the P--O bonds (IMS band).
The other IM band is observed in the SM calculation in the region $\approx 400$--700~cm$^{-1}$, and involves the O--P--O bending modes (IMB band) as shown in figure~\ref{fe_pe_dos.fig}.
The SM IMB band is associated with two peaks similarly to the AI band, 
but appears shifted $\approx 150$~cm$^{-1}$ towards higher frequencies
than the corresponding AI band. 
Unfortunately, any attempt to improve the fitting of these bands
by modifying different parameters led to a worsening of the overall fit.

Infrared measurements of the imaginary dielectric function
at 80~K in KDP with the electric field polarized along the FE axis show well-defined peaks up to $\approx 1400$~cm$^{-1}$~\cite{Sim88}.
The main resonances are: three in the region
(0--700)~cm$^{-1}$, centered at~$\approx 200$, 300 and 500~cm$^{-1}$,
and the other three in the region~(700--1400)~cm$^{-1}$, centered at $\approx 900$, 1000 and 1300~cm$^{-1}$.
These resonances also appear in the infrared spectrum obtained with the electric field polarized in the plane perpendicular to the FE axis~\cite{Sim88}.
Our SM phononic band structure in the FE phase is in close agreement
with the infrared spectra and enables us to identify the main experimental peaks. For instance, the experimental peaks centered at~$\approx 200$ and~1300~cm$^{-1}$
are in remarkable agreement with the SM lattice external (ET+ER) and HB bands centered
at~$\approx 150$ and 1300~cm$^{-1}$, respectively.
On the other hand, the bands observed experimentally at~$\approx 900$ and 1000~cm$^{-1}$ can be assigned
to our SM bands centered at $\approx 800$ and 1000~cm$^{-1}$
which are of IMS+HB character
(see the left-hand panel of figure~\ref{fe_pe_dos.fig}
and table~\ref{phonon1}).
The experimental peak observed at~$\approx 500$~cm$^{-1}$ should be related
to the SM IMB+HB band centered at nearly the
same frecuency, as shown in the left-hand panel of figure~\ref{fe_pe_dos.fig}.
These assignments coincide with those made
by the AI calculation of reference~\cite{Men18} 
(see also the similarities in the band distributions
for the SM and AI calculations in the left-hand panel of figure~\ref{fe_pe_dos.fig}).
On the other hand, the experimental peak at 300~cm$^{-1}$ can be ascribed to a mixed IMB+lattice
band centered at~$\approx 250$~cm$^{-1}$ 
observed in the SM and AI results which
mainly involve~O, K and P displacements (see the left-hand panel of figure~\ref{fe_pe_dos.fig}), or it can be 
related to the well-defined AI IMB band centered at $\approx 350$~cm$^{-1}$. Notice that in the latter case, the SM IMB band cannot account for it because
it is shifted towards higher frequencies.

\subsubsection{Effective Debye temperature}
\label{sec:3_2_2}

\begin{figure}[ht]
	\centerline{	\includegraphics[scale=0.35,angle=0]{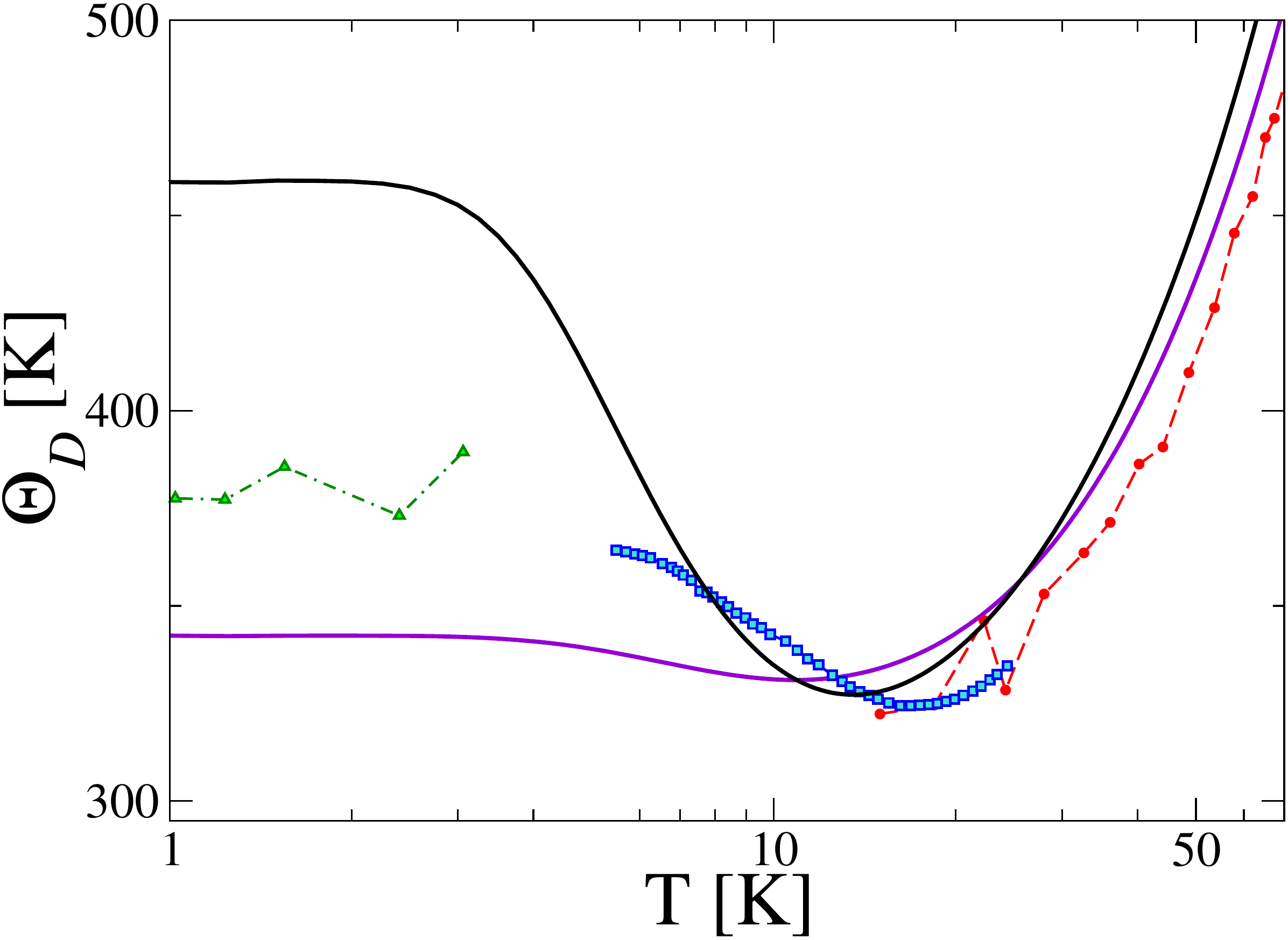}}
	\caption{(Colour online)  Equivalent Debye temperature $\Theta_{\rm D}(T)$ as a function of $T$ displayed in logarithmic scale. The theoretical results for the SM and AI calculations are shown in black and magenta solid lines, respectively. 		The experimental results  are plotted with red solid circles and dashed line (reference~\cite{Ste44}), with
		blue open squares (reference~\cite{Law82}), and with green open triangles (reference~\cite{Foo85}). }
	\label{debye.fig}
\end{figure}

With the aid of the developed model, we have calculated the effective (or equivalent) Debye temperature 
in the FE phase, $\Theta_{\rm D}(T)$, as an integral property of the total DOS~\cite{Bru82,Gie06,Las05b,Men18}. This temperature is associated to the value of the specific heat at each temperature~\cite{Bru82}.
The SM results for  $\Theta_{\rm D}(T)$ are plotted in figure~\ref{debye.fig} as a function of $T$ and
compared to the corresponding AI data of reference~\cite{Men18} and the available experimental data~\cite{Ste44,Law82,Foo85}.
A very good overall agreement of the SM results with the AI and experimental data is obtained, especially at temperatures $ \gtrsim 10$~K, as can be judged from figure~\ref{debye.fig}.

The low-temperature Debye region, where  $\Theta_{\rm D}(T)$ 
is  temperature independent and the Debye model is strictly valid, is observed
at $T \lesssim 3$~K and $T \lesssim 4$~K for both SM and AI calculations, respectively.
This is in qualitative agreement with the experimental Debye region observed
at $T \lesssim 5$~K considering the results of references~\cite{Law82,Foo85}, as shown in 
figure~\ref{debye.fig}.
As $T \rightarrow 0$, $\Theta_{\rm D}(T)$ takes the value of $\approx 450$~K for the SM
which is in qualitative agreement with the corresponding experimental and AI values, $\approx 380$~K and 
$\approx 350$~K, respectively.
The SM curve for $\Theta_{\rm D}(T)$ reaches a minimum value at $\approx 13$~K in very good 
agreement with the corresponding minima observed in the AI and experimental results,
as shown in figure~\ref{debye.fig}. However, the large drop observed in the SM result for $\Theta_{\rm D}(T)$ from the Debye region 
to the minimum value is overestimated in comparison with the corresponding
falls in the AI and experimental curves. The observed drop is a typical feature 
in the effective Debye-temperature profile which is produced 
by strong hybridizations of the acoustic and low-frequency optical branches towards the Brillouin zone boundary~\cite{Las05b}.

\subsection{Molecular dynamics simulations}
\label{sec:3_3}

Finally we have performed  classical-nuclei SMMD and AIMD simulations to study the phase transition in DKDP, which
is expected to have less quantum effects than KDP.
To track the transition we must first define the order parameter. To accomplish this, we assign to each proton a positive~($+$)
or negative~($-$) displacement $x_i$ from the instantaneous O--H--O bond center. 
The values of $x_i$ are considered
positive if the proton displacement direction in the particular O--H--O bond $i$
coincides with that observed for the global FE mode which causes polarization in the $z+$ direction, or negative otherwise~\cite{Men18}.
It is worth noting here that the coordinated proton
displacements around each phosphate following the
FE mode pattern are strongly correlated with the
local ionic plus electronic polarization along
the $z$ direction~\cite{Kov02,Kov05}. 
The order parameter $x$ is then computed as the spatial and time average of all proton displacements $x_i$. We 
define the critical temperature $T_c$ as the temperature at which
the order parameter falls to half of its maximum value in the ordered phase. 

\begin{figure}[ht]
\centerline{	\includegraphics[scale=0.35]{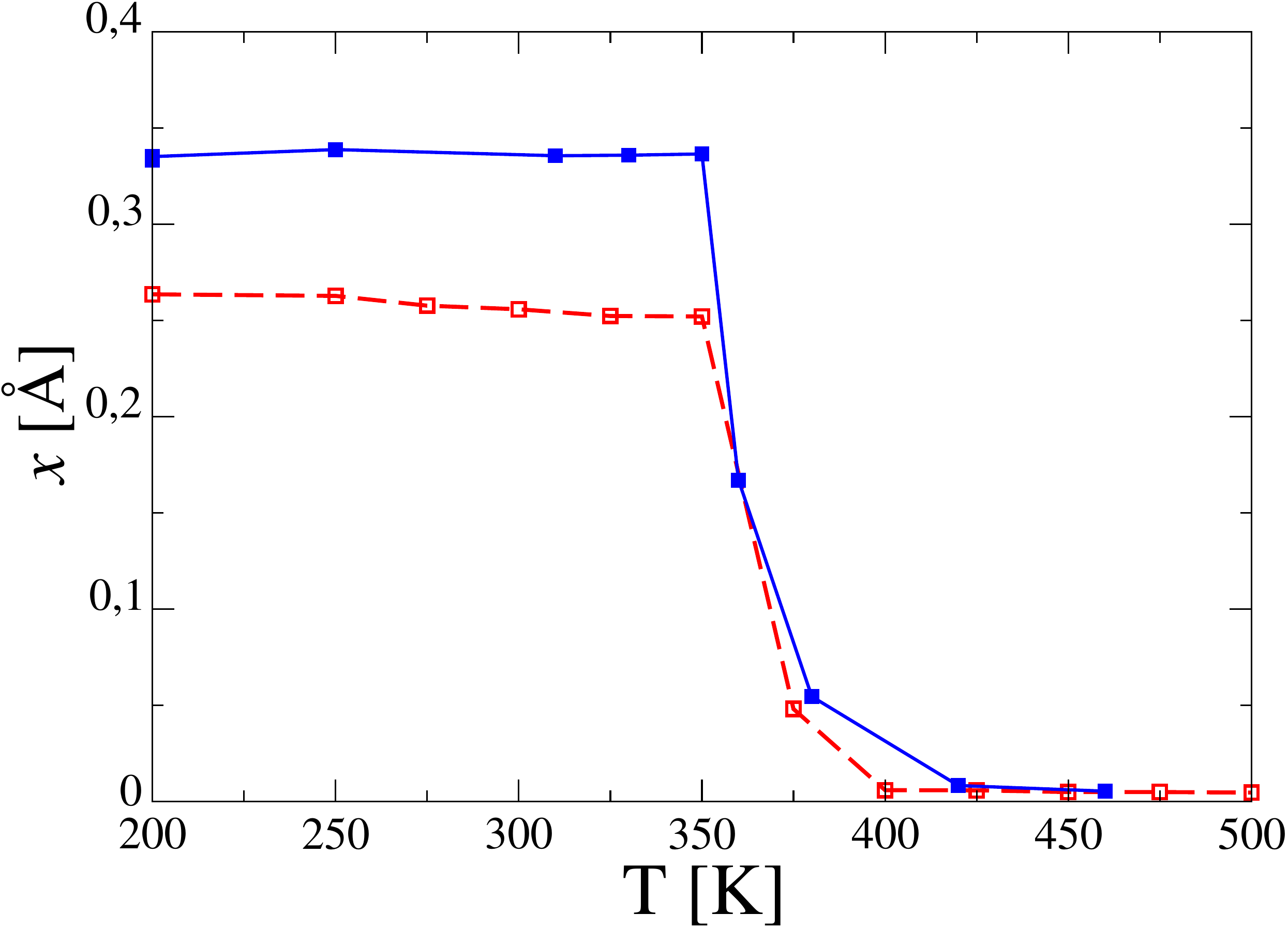}}
	\caption{(Colour online) Average order parameter $x$ as a function of $T$ obtained
		by AIMD and SMMD simulations. We show with red open squares and dashed lines the 
		AIMD results and with blue solid squares and solid lines the SMMD results for DKDP.
		Lines are guides to the eye only.
	}
	\label{mol_dyn.fig}
\end{figure}

Figure~\ref{mol_dyn.fig} depicts $x$ vs. $T$ for the AI and SM simulations. Simulations at different temperatures show that the system has a FE ordering up to a critical temperature $T_c$.
For larger temperatures, the average total  order parameter
vanishes and the PE phase arises.
We observe an excellent agreement for the values of $T_c$
obtained with the SMMD and AIMD simulations
for DKDP, $T_c^{\rm {SM}}$(DKDP)~$  \approx 360$~K
and $T_c^{\rm {AI}}$(DKDP)~$  \approx 365$~K, respectively.
Notice that the present molecular dynamics calculations do not consider
the quantum nature of the protons (and hence the possibility
of proton tunneling), and for this reason the
theoretical values of $T_c$ are expected to be larger than the corresponding
experimental value for DKDP, which is  $T_c^{\rm {exp}}$(DKDP)~$  \approx 229$~K~\cite{McM90,Nel91}.

\section{Summary}
\label{sec:4}

We have developed a SM fitted to AI results that include
non-local vdW corrections which are important to properly describe H-bond geometries
and global proton-transfer energy barriers. The development of this model is a first step for a future dynamical study in large-size systems of the elusive
 nature of the ferroelectric phase transition in KDP and other compounds of the H-bonded FE family. The SM was extensively tested 
by comparing structural and vibrational results to the corresponding
first principles and experimental data. The relaxed lattice and atomic structural parameters 
in the FE and PE phases are in very good correspondence with the respective vdW-corrected AI data 
as well as with the available experimental data. 

Regarding the vibrational properties, 
the SM $\Gamma$-point phonons obtained for the FE and PE phases are in good overall agreement 
with the corresponding AI results and the available experimental data.
The FE  mode at the Brillouin zone center calculated
with the developed SM is unstable in the PE phase, in agreement with the AI results.
The total SM-DOSs in the FE and PE phases
are also in very good accordance with the corresponding
AI data suggesting that the model represents well the vibrational properties of KDP throughout the whole Brillouin zone. 
On the other hand, the obtained SM IMB band is in qualitative agreement with the corresponding AI band, but is shifted 
$\approx 150$~cm$^{-1}$ towards higher frequencies.
The SM and AI bands in the FE phase are also in very good agreement with the resonances
observed in infrared measurements of the imaginary
dielectric function at 80~K in KDP, which enables us to associate
the measured peaks with the corresponding phononic displacement
patterns. 
Using the phonon DOS obtained in the FE phase
with the developed SM, we have
computed the effective Debye temperature which is in very good agreement with the corresponding AI and experimental data.
Finally, we have carried out SMMD simulations for classical nuclei and found a FE--PE phase transition for DKDP at $\approx 360$~K
which is in very good agreement with the corresponding AIMD simulations  
that include nonlocal dispersion effects via the vdW-DF approach. 

\section*{Acknowledgements}

We acknowledge support from Consejo Nacional de
Investigaciones Cient\'ificas y T\'ecnicas (CONICET) and Centro de Simulaci\'on
Computacional para Aplicaciones Tecnol\'ogicas (CSC-CONICET) for
the computing hours provided to perform the simulations of this work.

\ukrainianpart

\title{Оболонкова модель та першопринципні розрахунки коливних, структурних і сегнетоелектричних властивостей KH$_2$PO$_4$}
\author{Р. Е. Менхон, Ф. Торрезі, Х. Ласаве, С. Коваль}
\address{
	Інститут фізики Росаріо, Національний університет Росаріо та Національна рада з науково-технічних досліджень, вул. 27 лютого, 210 Bis, 2000 Росаріо, Аргентина 
}

\makeukrtitle

\begin{abstract}
	\tolerance=3000%
	Розвинено оболонкову модель для	дигідрофосфату калію (KDP), яку прив'язано до результатів
	першопринципних ({\it ab initio}) обчислень, що враховують нелокальні поправки ван дер Ваальса. Модель ретельно дос\-ліджується шляхом порівняння
	результатів для структурних, коливних і сегнетоелектричних характерис\-тик з результатами першопринципних розрахунків та експериментальними даними. Релаксаційні струк\-турні параметри дуже добре узгоджуються з результатами {\it ab initio} та з наявними експериментальними даними. Повна густина фононів та їх густина в $\Gamma$ точці зони Бріллюена у сегнетоелектричній та параелек\-тричній фазах, розраховані в рамках оболонкової моделі, загалом добре узгоджуються з відповідними результатaми {\it ab initio} та експериментальними даними. Також обчислено ефективну температуру Дебая як функцію $T$, яка добре відповідає результатам  {\it ab initio} та експериментальним даним. Температура фазового переходу ``сегнетофаза-парафаза'', отримана шляхом класичної молекулярної динаміки для розвиненої оболонкової моделі, складає $\approx 360$~K, що чудово узгоджується з результатами першопринципних молекулярно-динамічних розрахунків.
	\keywords сегнетоелектрики, водневі зв'язки, фазові переходи, оболонкова модель, метод функціоналу густини
	
\end{abstract}

 \lastpage
 \end{document}